\documentclass[twocolumn,showpacs,preprintnumbers,amsmath,amssymb]{revtex4}
\usepackage{graphics}
\begin{document}
\title{EqRank: Theme Evolution in Citation Graphs}

\author{G. B. Pivovarov}
\email{gbpivo@ms2.inr.ac.ru}
\affiliation{Institute for Nuclear Research,\\
Moscow, 117312 Russia}

\author{S. E. Trunov}
\email{trunov@msk.seus.ru}
\affiliation{SEUS, Moscow, Russia}

\date{October 12, 2005}
 
\begin{abstract}
Time evolution of the classification scheme generated by the EqRank algorithm is studied with hep-th citation graph
as an example. Intuitive expectations about evolution of an adequate classification scheme for a growing set of 
objects are formulated. Evolution compliant with these expectations is called natural. It is demonstrated that EqRank
yields a naturally evolving classification scheme. We conclude that EqRank can be used as a means to detect new scientific themes, and to track their development.
\end{abstract}


\maketitle

\section{Introduction}

Detection of ''communities'' has lately became a popular subject in the studies of complex linked networks \cite{1}.
For a citation graph, communities are identified as scientific themes. Most of the papers devoted to this subject deal with separate snapshots of graphs disregarding their time evolution \cite{2}.

If we consider the time evolution of a graph evolving nonrandomly, we expect to see regularities in the evolution. Examples are citation graphs or the Web graph (for the latter, vertexes are web pages, and edges are the hyper-links). 
For citation graphs of scientific papers in a subject field (vertexes of the graph are the papers, edges, the citations), communities are themes forming a classification scheme of the subject field \cite{3,4}. As time goes, the classification scheme may undergo changes, but the changes should be of a regular character. There should exist a relatively large stable set of core themes that does not change. A few themes may disappear (their contents absorbed by other themes), and new themes may appear. 

If we use a classification algorithm, and claim that it yields adequate themes, we should check if its outcome generates a dynamics of graph partitions with the above properties. In the paper, we present a modification of the EqRank algorithm \cite{8} that passes such a check. 

The paper is organized as follows: In Section 2, we briefly review previous work on the problem. In Section 3, we formalize the requirements on the dynamics of partitions outlined above. In Section 4, we outline the EqRank algorithm. Section 5 presents results of applications of the algorithm to the hep-th citation graph.

\section{Related Work}

Despite the forty year history of citation analysis \cite{5}, the problems listed above have only recently become a subject of constant interest. In \cite{6}, time evolution of themes in computer science has been analysed. The analysis was based on application of a dedicated graph clustering algorithm applied to the data of Citeseer library. 
In the analysis, a fixed list of themes was used, which is not a consistent approach to the problem. Approach of \cite{7} is closer in spirit to our paper. It was observed in this paper that hierarchical agglomerative clustering yields an
unstable partition, and that there exists a small subset of clusters that remains stable against random perturbations of the graph. Namely these stable clusters were called natural communities, and were identified as scientific themes. In our approach, we also consider stability of the partition as a major requirement. Comparing our approach to the one of \cite{7}, we point out two distinctions.

First, we concentrate not on the stability of particular clusters, but on the stability of the whole partition. As pointed out in \cite{7}, there exist many algorithms yielding partitions with a stable core. On the other hand, there must be much less algorithms yielding partitions that are stable as a whole. We describe below a modification of the EqRank algorithm possessing namely this property.

Secondly, we require not only the stability of partitions against small random perturbations, but also against large regular perturbations, among which we consider in particular the time evolution of graphs appearing in actual applications. For this latter case, we transform the stability requirement to requirements of {\it natural dynamics}. These requirements are naturally expected to be satisfied by the dynamics of evolving partitions.

\section{Natural Theme Dynamics}

We consider a graph $G(t)$ growing in time. Take two subsequent moments of time, $t_2 > t_1$. As the graph is growing, 
$G(t_1)$ is a subgraph of $G(t_2)$. Therefore, any partition of $G(t_2)$ induces a partition of $G(t_1)$. Thus, for a fixed classification algorithm, we have two partitions of the set of vetexes ${\mathbf T}$ of the graph $G(t_1)$:
\begin{eqnarray}
{\mathbf T_1} &=& \{T_1,...,T_n\},\nonumber\\
{\mathbf T_2} &=& \{T_{1'} ,...,T_{k'}\},\nonumber
\end{eqnarray}
where $T_i$ and $T_{j'}$ are the clusters at two moments of time. We expect that ${\mathbf T_2}$ is ${\mathbf T_1}$ with new 
clusters (themes) added, and some themes removed:
\begin{equation}
\label{t2}
{\mathbf T_2} = {\mathbf T_1}\cup {\mathbf N\mathbf{T}}\backslash{\mathbf A\mathbf{T}_1},
\end{equation}
where ${\mathbf N\mathbf{T}}$ is the set of new themes, and ${\mathbf A\mathbf T_1}$ is the set of absorbed themes. 
To give a meaning to this equality, we have to set a mapping between the elements of $\mathbf T_2$ and $\mathbf T_1$:
\begin{equation}
Map_1(T_i) = T_{j'},
\end{equation}
where $T_{j'}$ is the cluster of the second partition that has a maximal intersection with $T_i$.
Analogously, we define $Map_2$ relating a cluster of the second partition to a cluster of the first
(notice that generally $Map_1\neq Map_2$). We can now define the subset of {\it stable themes} as a subset of clusters
of the first partition satisfying the following requirement:
\begin{equation}
\mathbf S\mathbf T_1 = \{T_i: Map_2\circ Map_1(T_i)=T_i\}.
\end{equation}

On the subset of stable themes, we can consider time development of a theme: if $T\in\mathbf S\mathbf T$, $Map_1(T)$ is
$T$ at a subsequent moment of time. 

Equation (\ref{t2}) can be rewritten as follows:
\begin{eqnarray}
\mathbf T_2 &=& Map_1(\mathbf S\mathbf T_1)\cup \mathbf N\mathbf T,\\
\mathbf T_1 &=& \mathbf S\mathbf T_1\cup \mathbf A\mathbf T_1.
\end{eqnarray}
The sets $\mathbf N\mathbf T$ and $\mathbf A\mathbf T_1$ are respectively complements to the sets $Map_1(\mathbf S\mathbf T_1)$ and $\mathbf S\mathbf T_1$. For $T'\in \mathbf N\mathbf T$, we say that $T'$ has broke away from a cluster $T\in \mathbf T_1$
if $Map_2(T')=T$. For $T\in\mathbf A\mathbf T$, we say that $T$ has been absorbed by $T'\in\mathbf T_2$ if $Map_1(T)=T'$.
 
The above characterizations can be applied to any pair of partitions of a graph. Practically, an adequate clustering
algorithm applied to an evolving graph should yield partitions satisfying a number of requirements.

The first requirement is that the initial partition $\mathbf T_1$ would mostly consist of the stable subset 
$\mathbf S\mathbf T$, and the fraction of papers belonging to the stable subset of the partition would be high. We write it as the {\it classification scheme stability requirements}:
\begin{eqnarray}
\label{cssr1}
|\mathbf S\mathbf T_1|/|\mathbf T_1| \approx 1,\\
\label{cssr2}
|\cup T_j: T_j\in \mathbf S\mathbf T_1|/|\cup T_j\in \mathbf T_1| \approx 1.
\end{eqnarray}

Notice that we do not require that $|\mathbf S\mathbf T|/|\mathbf T_2|\approx 1$, because we want to keep in the consideration the cases than the number of new themes is comparable to the number stable themes. However, we require that the number of old papers belonging to new themes would not be large (see (\ref{cssr2})), i.e., new themes should mostly consist of new papers.

Classification scheme stability requirements (\ref{cssr1}),(\ref{cssr2}) restrict evolution of the partition on the level of themes. We also need to restrict the way specific papers move around the partition. We call this set of requirements {\it indexing consistency requirements}. They read as follows:
\begin{itemize}
\item
If a paper belongs to a stable theme in the partition $\mathbf T_1$, it either remains in the same theme in the partition $\mathbf T_2$, or moves to a new theme that breaks away from the theme in the first partition.
\item
If a paper belongs to a new theme in the partition $\mathbf T_2$, it belongs to the stable theme of $\mathbf T_1$ from which the new theme has broken away.
\item
If a paper belongs to a theme of the partition $\mathbf T_1$ that gets absorbed by a stable theme, it belongs to the stable theme of the partition $\mathbf T_2$.
\end{itemize}

The indexing consistency requirements imply the following scenario for forming the new partition from the old one:
\begin{itemize}
\item
Select the subset $\mathbf S\mathbf T_1$ of stable clusters.
\item
For each cluster $S\in\mathbf S\mathbf T_1$, chip off new clusters. The remaining part of $S$ becomes the stable cluster $S'$ of the partition $\mathbf T_2$.
\item
Each of the rest of the clusters of $\mathbf T_1$ is absorbed by a particular stable cluster of $\mathbf T_2$.
\end{itemize}
 
Evidently, this is a very specific scenario. Let us reiterate it less formally: Evolution consists in emanation of new themes from stable themes (several new themes may chip off from a stable theme), and in absorption of unstable themes by stable themes. A separate feature that should be stressed is that a new theme has a unique parent stable theme (in a more general scenario, a new theme could be formed from pieces originating from different stable themes), and that
an unstable theme is absorbed by a single stable theme (in a more general scenario, it could be redistributed among several stable themes). Summarizing all the restrictions, we require that evolution of the partition would keep identity of themes (reshuffling of splinters of clusters into new clusters is forbidden).

We call a dynamics of partition the natural dynamics if it satisfies the above classification scheme stability and index consistency requirements.

In practice the above requirements can be satisfied only approximately. We introduce three coefficients to quantify the deviation of the real dynamics from a natural dynamics:
\begin{itemize}
\item
$CSC_1$ (the first classification stability coefficient) is the percentage of the stable themes with respect to the total number of themes.
\item
$CSC_2$ (the second classification stability coefficient) is the percentage of papers in the stable themes with respect to the total number of papers.
\item
$TMC$ (theme mixing coefficient) is the percentage of the papers violating the indexing consistency requirements.
\end{itemize} 

\section{EqRank as Algorithm of Self-Organization}

The EqRank algorithm was presented in \cite{8}. Here we give an informal description using a metaphor of forming coalitions in a social network, and present a modification of the algorithm yielding partitions with improved stability properties. In particular, the modified algorithm leads to satisfactory numerical values of the above coefficients, characterizing the closeness of the evolution to the natural one.

Consider a social network with vertexes representing persons, and links, the trust or sympathy to the persons to which they are pointing at. The links are weighted, the weight is measuring the extent of trust. 
Let there be a reason forcing people to form coalitions, and let the decision to join a coalition be taken on
the basis of trust to the nearest neighbors.

We define the rule of joining the coalition recursively: a person joins the coalition joined by the person enjoying the maximal trust of the first person. There are many partitions satisfying this rule. In \cite{8} we demonstrated that EqRnk yields the maximally detailed partition satisfying this rule. This means that any other partition satisfying the above rule is a coarse-graining of the partition yielded by EqRank. 

EqRank follows the practice of forming coalitions: a person joins the person whom she trusts the most, and brings
to the coalition the persons trusting her the most. The latter also bring the persons trusting them the most, 
and so on. Programming this process is simple: The algorithm starts from discarding nonmaximal links. 
In this way the maximal graph is formed containing only the links expressing the relations of maximal trust. 
(Note that the maximal graph is directed: If a person trusts the most a person, the latter may
trust the most a different person.) Because the maximal link is unique (this is a simplifying assumption), each vertex of the maximal graph 
defines unambigously a chain that starts form the vertex, and formed from the vertexes joined by the maximal
links. Because the graph is finite, each chain ends on a cycle, which is a nontrivial strongly connected 
component of the maximal graph. By the definition, all the vertexes of the chain belong to the same 
cluster (coalition). And the chains that end up on the same cycle should also belong to the same coalition.

The EqRank pseudocode is reduced to the following operations:
\begin{itemize}
\item
Selection of the maximal subgraph.
\item
Computation of the strongly connected components of the maximal subgraph, and contraction of
the strongly connected componenets to the vertexes of a factorgraph.
\item
Selection in the factorgraph of the vertexes without outgoing links (final vertexes). 
The cycles of the initial graph correspond to the final vertexes.
\item
Forming a coalition from all the vertexes of the factorgraph from which it is possible 
to reach one and the same final vertex. 
\end{itemize}

Let us consider again our methafor of forming coalitions in a social network. We can compute
{\it total trust} of a person to arbitrary coalition as the sum of the link weights running
over all the links joining the person with the members of the coalition. We say that the coalition 
a person belongs to {\it meets her expectations} if the total trust of the person to her own
coalition is larger than her total trust to any other coalition. We call a coalition proper
if it meets expectations of each of its members. Condition defining proper coalition is 
much weaker than the one defining {\it communinty in the strong sense} \cite{1} (the latter
require that the total trust of a person to her coalition would be grater than the sum of total
trusts to all other coalitions). The condition defining the proper coalition allows us to start 
a process of coalition restructuring. Namely, we let each person to go over to the coalition
she trusts the most. After all the persons performe the transition, each of them compures again the
trust to the new coalitions. It may happen that the coalition a person finds herself in after the 
first transition again does not meet her expectations, becaus it may happen that the most trusted
members of the old coalition have left to new coalitions. In this case, the restructuring of the 
coalitions repeats itself. Assuming that expectations of most of the persons are met
with the initial coalitions, we expect that the restructured coalitions meet the expectations of 
all the persons after finite number of restructuring iterations. We note that the number
of improved coalitions is generally reduced with respect to the number of intitial coalitions, 
because initially there may exist coalitions for which all the members leave to another coalitions.

We now give a formal definition of the above iterations. Let ${\mathbf T}=\{T_1,...,T_n\}$ 
be a partition of the
weighted graph $G=(V,E,w)$. Partition ${\mathbf T}$ is defined by the function $t(x)=i$, where $x\in T_i$. Let 
us define the function $D(V,{\mathbf T})$: $D(x,T_i)=\sum_{y\in T_i}w(x,y)$. This function quantifies 
the closeness 
between the vertex $x$ and the cluster $T_i$. Let us define the sequence of partitions ${\mathbf T_n}$ as follows:
\begin{equation}
t_{n+1}(x)=j,
\end{equation}
where $T_j$ belongs to ${\mathbf T_n}$ and gives maximum to the function $D(x,{\mathbf T_n})$. 
The process starts from
$t_0\equiv t$. In words, the partition ${\mathbf T_{n+1}}$ indexes $x$ with the index $j$ of the cluster
$T_j$ of the partition ${\mathbf T_n}$ that is the closest one to $x$.
We say that the sequence converges at the vertex $x$ if $t_{n+1}(x)=t_n(x)$ starting from
some $n$, which may depend on $x$. 

Let $V_{max}({\mathbf T})$ be the maximal subset of the graph vertexes on which the sequence 
${\mathbf T_n}$ is converging. Practically, a good clustering algorithm should yield 
a partition ${\mathbf T}$ for which $V_{max}({\mathbf T})$ is large enough, and the convergence is
fast. We denote the limiting partition of $V_{max}({\mathbf T})$ as $lim({\mathbf T})$, and call the
above restructuring process the reindexing process.

\begin{figure}
\label{reindex}
    \begin{center}
      \resizebox{80mm}{!}{\includegraphics{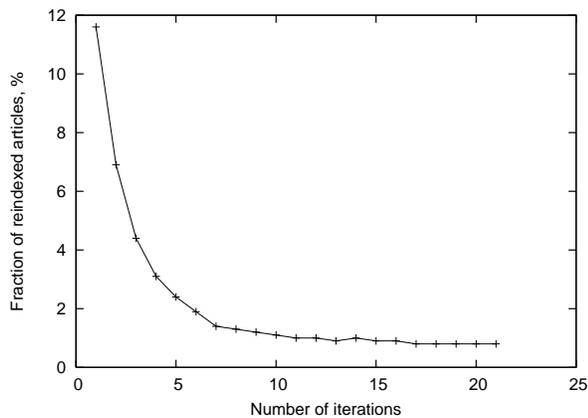}}
      \caption{Iterations decrease reindexing}
    \end{center}
  \end{figure}

We suggest the following modification of the EqRank algorithm:
\begin{itemize}
\item
The initial partition is yielded by the standard EqRank algorithm (see
the above pseudocode).
\item
Sarting form the initial partition, the limiting partition is constructed by the
above reindexing process. The limiting partition is taken as the result yielded
by the modified EqRank.
\end{itemize}

Fig. 1 shows convergence of reindexing for the initial partition yielded by
the standard EqRank applied to the hep-th citation graph. As seen, after 10 iterations,
the reindexing process converges for 99\% of the graph vertexes. 

\begin{figure}
    \begin{center}
      \resizebox{80mm}{!}{\includegraphics{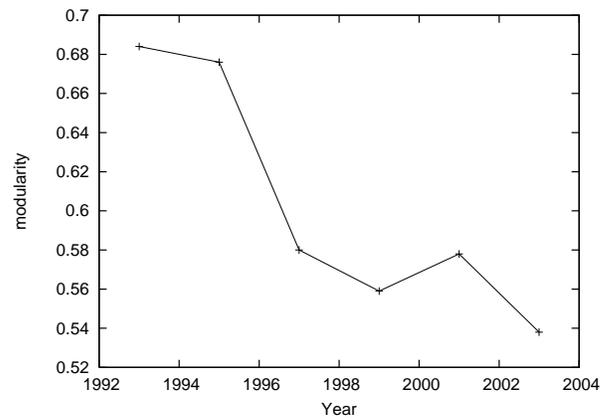}}
      \caption{Modularity from 1993 to 2003}
    \end{center}
  \end{figure}

 \begin{figure}
    \begin{center}
      \resizebox{80mm}{!}{\includegraphics{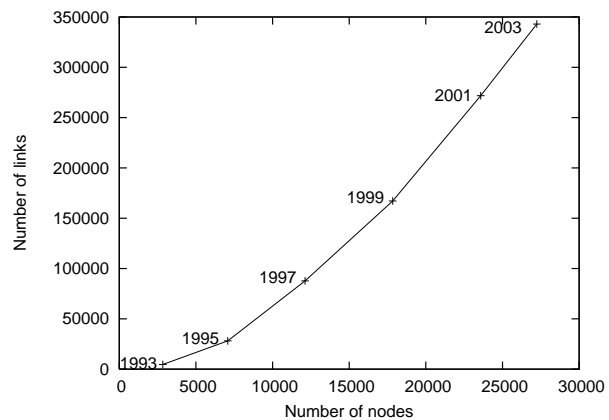}}
      \caption{Growth of node and link numbers from 1993 to 2003}
    \end{center}
  \end{figure}
\begin{figure}
    \begin{center}
      \resizebox{80mm}{!}{\includegraphics{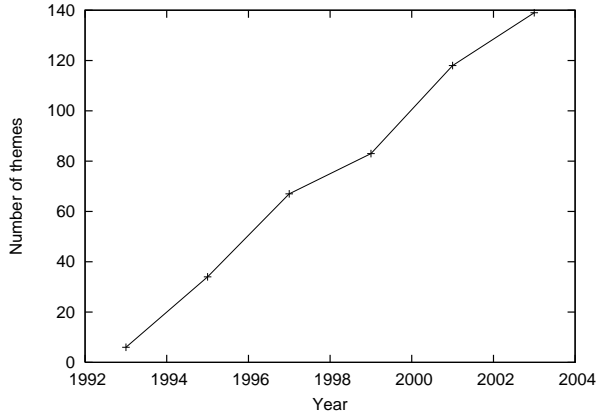}}
      \caption{Number of themes from 1993 to 2003}
    \end{center}
  \end{figure}

\begin{figure}
    \begin{center}
      \resizebox{80mm}{!}{\includegraphics{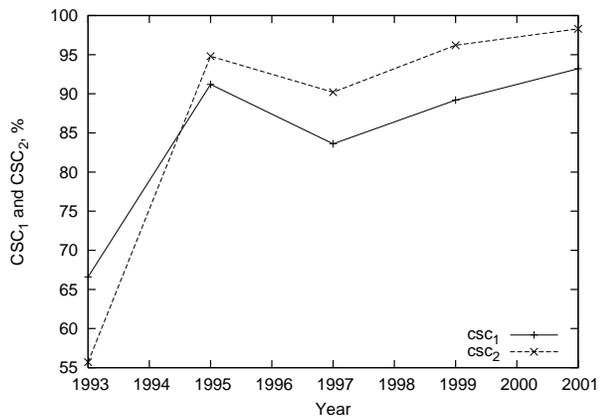}}
      \caption{Classification stability coefficients from 1993 to 2001}
    \end{center}
  \end{figure}

\begin{figure}
    \begin{center}
      \resizebox{80mm}{!}{\includegraphics{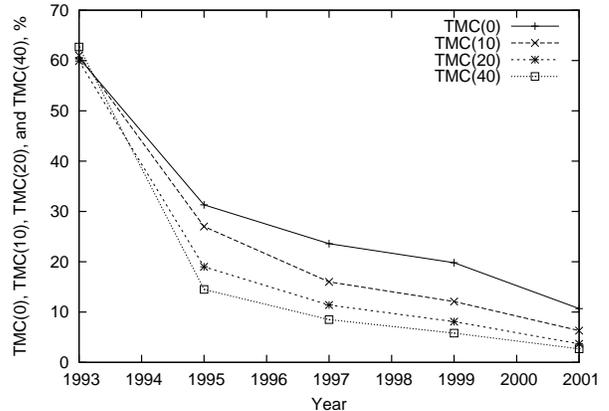}}
      \caption{Theme mixing coefficients from 1993 to 2001}
    \end{center}
  \end{figure}

\begin{figure}
    \begin{center}
      \resizebox{80mm}{!}{\includegraphics{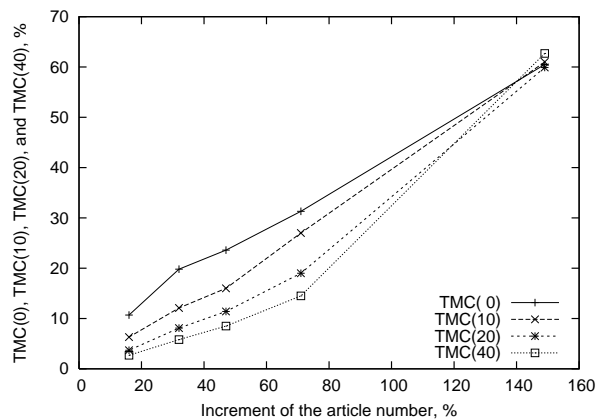}}
      \caption{Theme mixing coefficient}
    \end{center}
  \end{figure}

We conclude this section with the following observations:
\begin{itemize}
\item
Selection of the maximal subgraph is an important stage of the algorithm.
It may seem that this selection discards substantial information on the
graph encoded in the nonmaximal links. This is not the case if a
graph-based proximity measure is used to determine the link weights. In our experiments,
we used a linear combination of the co-citation and bibliographic coupling \cite{5}. 
In this case, even the maximal links encode information on the global network topology.
Fig. 2 can be used to support this assertion. It shows the modularity of the partitions yeilded
by EqRank applied to six snapshots of the hep-th citation graph. The modularity was introduced 
in \cite{9}. It quantifies both the hidden graph community structure and the quality of an
algorithm aimed at approximating this structure. Computation of mudularity involves all the links
regardless of their weights. The graphs with community structure should have modularity in the range from
0.3 to 0.7. The plot shows that the EqRank partitions have the modularity in this range.
\item
The reidexing process improves the quality of the community structure implied by the limiting 
partition $lim({\mathbf T})$. It is even more important that evolution of the limiting partition
is closer to a natural one (see the above definition of natural evolution) than the one
of the initial partition. This is an experimental fact (see below): the values of the coefficients
quantifying closeness of the evolution to a natural one are better for the limiting partition
than they are for the initial one.
\end{itemize}

\section{Hep-th Theme Dynamics}

We used hep-th citation graph \cite{10} to study the dynamics of the partitions yielded by the 
EqRank algorithm. The graph we have studied contained the information on the citations in the
papers from the hep-th sector of the electronic archive http://arxiv.org for the years 1992--2003.

We have taken six snapshots of this graph, at the end of 1993, 1995, 1997, 1999, 2001, and 2003.
(This means that the first snapshot contains all the papers appeared before 1994, the second, before 1996, etc.)
Fig. 3 shows the growth of the number of the graph vertexes (nodes) and links during these years. 
In contrast to our previous experiment with hep-th \cite{8}, we used undirected version of the citation
graph. The links of the graph were weighted with a combination of co-citation and bibliographic 
coupling:
\begin{equation}
W(x,y)=a A^T A +(1-a)A A^T,
\end{equation}
where $a=0.9$ and $A$ is the adjacency matrix of the graph. For each graph, we applied the modified
EqRank. The total number of themes have grown in this period from 6 (in 1993) to 139 (in 2003) (see Fig. 4).

To make an estimate of the closeness of the partition dynamics to a natural one, we computed the first and
the second
classification stability coefficients (Fig. 5). The plot shows that after 1995 the values of
this coefficient are above 80\%, which we consider as satisfactory values. Next we computed the theme
mixing coefficient, TMC (Fig. 6). Again, after 1995, this coefficient ranges from 10 to 30\%. 
We point out that the number of papers in hep-th was growing fast, and, in view of this, 
the above values of TMC do not seem to be high. The plot on Fig. 7 demonstrates that TMC grows
with the increment in the number of papers. For example, the number of vertexes have grown by 16\% from 2001 to 2003, 
and TMC was about 10\%. Apart from TMC defined above, we computed also TMC(cut). In computing it, 
only the papers whose citation index exceeds the cut were taken into account. The values of TMC(cut)
are informative, because in practice it is important to give a correct indexing
namely to the well cited papers that can be viewed as specific theme attributes. The higher is the
citation index of a paper, the more important is that its theme index would behave properly. The
plots on Figs. 6 and 7 demonstrate the desired behavior of TMC depending on the value of the cut.
The higher is the cut, the lower is the value of the corresponding TMC. For example, TMC(40) was about
3\% in 2001--2003, while TMC without cut was 10\%.

Anomalous values of the coefficients in 1993--1995 can be explained as follows. In this period, the archive
was filled fast with the themes already existed outside the archive, and the classification 
index was changing dramatically. After 1995, the archive became the standard location for 
the papers in its subject field, most of the themes were represented relatively well, and 
the coefficients relaxed to satisfactory values.

 

\begin{table*}
\caption{Comparison}
\begin{ruledtabular}
\begin{tabular}{|l|l|l|l|l|l|}
   \hline
Number of   & Theme keywords & Authority papers ({\it italic} marks    & Theme's  & Peskin's&Context\\
articles    &                & the papers mentioned in Peskin's review)& years    & year    &       \\ 
   \hline
4338        & string theory;  &{\it M Theory As A Matrix Model:}   &     1993---   &  1997  &Mentioned in the\\
            &heterotic string;&{\it A Conjecture [hep-th/9610043];}&     1995      &        &Section on \\
            &type iib;        &{\it Dirichlet-Branes and}          &               &        &M-theory\\
            &type iia         &{\it Ramond-Ramond Charges }&      &        &\\
            &                 &{\it [hep-th/9510017]; }    &               &        &\\
            &                 &{\it String Theory Dynamics In}&           &        &\\
            &&{\it Various Dimensions [hep-th/9503124]}&&&\\
   \hline 
3008        &black hole; &{\it The Large N Limit of Superconformal}&    1997---    & 1998   &''Maldacena's \\     
            &ads/cft ñorrespondence;&{\it Field Theories and }     &    1999       &        & correspondence...''\\
            &gauged supergravity;&{\it Supergravity [hep-th/9711200];} &               &        &\\
            &higher spin&{\it Anti De Sitter Space}      &               &        &\\
            &&{\it And Holography [hep-th/9802150];}   &               &        &\\
            &&{\it Gauge Theory Correlators}           &               &        &\\
            &&{\it from Non-Critical String Theory}               &               &        &\\
            &&{\it [hep-th/9802109];}                               &               &        &\\
            &&{\it Anti-de Sitter Space,}                         &               &        &\\
            &&{\it Thermal Phase Transition,}                     &               &        &\\
            &&{\it And Confinement In Gauge Theories}&&&\\
            &&{\it  [hep-th/9803131];}&&&\\
            &&{\it 4d Conformal Field Theories}&&&\\
            &&{\it and Strings on Orbifolds [hep-th/9802183]}&&&\\
            \hline
1443 &n=2 supersymmetric;&{\it Monopole Condensation, And Confinement}&1993---&1997&''Witten uses brane\\
&integrable system;&{\it In N=2 Supersymmetric Yang-Mills}&1995&&dynamics to give a new\\
&supersymmetric gauge;&{\it Theory [hep-th/9407087];}&&&derivation of the ...\\
&lax pair&{\it Monopoles, Duality and Chiral }&&&exact solutions of four-\\
&&{\it Symmetry Breaking in N=2 }&&&dimensional N=2 super-\\
&&{\it Supersymmetric QCD [hep-th/9408099];}&&&Yang-Mills theory of\\
&&{\it Electric-Magnetic Duality in}&&&Seiberg and Witten...\\
&&{\it Supersymmetric Non-Abelian}&&&Seiberg's ...picture\\
&&{\it Gauge Theories  [hep-th/9411149]}&&&of N=1 super-Yang-Mills\\
&&&&& theory can also be \\
&&&&& understood in this way...''\\
            \hline
1398&noncommutative field;&{\it Noncommutative Geometry and}&1997---&2000&Mentioned in the\\
&seiberg-witten map;&{\it Matrix Theory: Compactification}&1999&&Section on \\
&open string;&{\it on Tori [hep-th/9711162];}&&&Noncommutative\\
&noncommutative &{\it D-branes and the Noncommutative}&&&Geometry\\
&geometry&{\it Torus [hep-th/9711165];}&&&\\
&&{\it String Theory and}&&&\\
&&{\it Noncommutative Geometry}&&&\\
&&{\it [hep-th/9908142];}&&&\\
&&{\it Noncommutative Perturbative}&&&\\
&&{\it Dynamics [hep-th/9912072]}&&&\\
           \hline
1185&brane world;&{\it An Alternative to Compactification }&1997---&2000&Mentioned in the\\
&cosmological constant;&{\it [hep-th/9906064];}&1999&&Section on\\
&extra dimension;&{\it Modeling the Fifth Dimension}&&&Extra Space Dimensions\\
&randall-sundrum&{\it with Scalars and Gravity}&&&\\
&model&{\it [hep-th/9909134];}&&&\\
&&{\it On Conventional Cosmology}&&&\\
&&{\it from a Brane Universe}&&&\\
&&{\it [hep-th/9905012]}&&&\\
           \hline
\end{tabular}
\end{ruledtabular}
\end{table*} 
\begin{table*}
\caption{Comparison (continuation)}
\begin{ruledtabular}
\begin{tabular}{|l|l|l|l|l|l|}
   \hline 
Number of   & Theme keywords & Authority papers ({\it italic} marks    & Theme's  & Peskin's&Context\\
articles    &                & the papers mentioned in Peskin's review)& years    & year    &       \\ 
   \hline           
437&extremal black;&{\it Microscopic Origin of the}&1995---&1997&''Strominger and Vafa\\
&greybody factor;&{\it Bekenstein-Hawking Entropy}&1997&& and Callan and  \\
&null model;&{\it [hep-th/9601029];}&&&Maldacena have shown  \\
&bekenstein-hawking&{\it D-brane Approach to Black}&&&that black hole  \\
&formula&{\it Hole Quantum Mechanics}&&&solutions of string \\
&&{\it [hep-th/9602043]}&&&theory can be thought \\
&&&&&to contain D-branes in  \\
&&&&&their compactified \\
&&&&&dimensions...''\\ 
           \hline
383&local brst;&Antibracket, Antifields and Gauge&1991---&&\\
&consistent deformation;&Theory Quantization &1993&&\\
&nonlocal regularization;&[hep-th/9412228];&&&\\
&brst cohomology;&Local BRST Cohomology in the&&&\\
&&Antifield Formalism: I. General&&&\\
&&Theorems [hep-th/9405109];&&&\\
           \hline
352&lambda model;&Two Dimensional QCD is a String&1993---&&\\
&wilson loop;&Theory [hep-th/9301068];&1995&&\\
&vapour phase;&Two Dimensional Gauge Theories&&&\\
&two-dimensional&Revisited [hep-th/9204083];&&&\\
&yang-mills&Twists and Wilson Loops&&&\\
&&in the String Theory of Two&&&\\
&&Dimensional QCD [hep-th/9303046]&&&\\
          \hline
321&integrable boundary;&Boundary S-Matrix and Boundary&1995---&&\\
&s matrix;&State in Two-Dimensional Integrable&1997&&\\
&bethe ansatz;&Quantum Field Theory&&&\\
&affine toda&[hep-th/9306002];&&&\\
&&Factorized Scattering in the&&&\\
&&Presence of Reflecting Boundaries&&&\\
&&[hep-th/9304141]&&&\\ 
         \hline
303&dilaton gravity;&The Stretched Horizon and&1991---&&\\
&hawking radiation;&Black Hole Complementarity&1993&&\\
&hole evaporation;&[hep-th/9306069];&&&\\
&two-dimensional&The Endpoint of Hawking&&&\\
&dilaton&Evaporation [hep-th9206070]&&&\\
        \hline
289&penrose limit;&{\it Strings in Flat Space and pp}&2001---&2003&''The theory of strings on\\
&pp-wave background;&{\it waves from} ${\cal N}=4$&2003&&gravitation wave\\
&plane wave;&{\it Super Yang Mills} &&&backgrounds...''\\
&bmn operator&{\it [hep-th/0202021]}&&&\\
        \hline
275&gaugino ñondensation;&{\it Type IIB Superstrings, BPS Monopoles,}&1999---&1997&''This suggests that there is\\
&heterotic m-theory;&{\it And Three-Dimensional Gauge}&2001&&a connection between the\\
&supersymmetry &{\it Dynamics [hep-th/9611230];}&&&geometry of branes and the \\
&breaking&{\it Solutions Of Four-Dimensional}&&&exact properties of these \\
&&{\it Field Theories via}&&&gauge theories...''\\
&&{\it M Theory [hep-th/9710136]}&&&\\
        \hline 
110&rolling tachyon;&{\it Rolling Tachyon}&2001---&2003&''...study of the explicit\\
&tachyon matter;&{\it [hep-th/0203211]}&2003&&time-dependent evolution \\
&tachyon field&&&&of unstable brane \\
&&&&&configurations...''\\
\end{tabular}
\end{ruledtabular}
\end{table*}

Tables 1 and 2 of the Appendix represent the themes of the classification obtained for the hep-th
citation graph as it was in the mid of 2003. Due to the lack of space, we present only the upper themes
ordered by the number of papers in the theme. Most of the themes are easily recognised. For comparison,
we used the annual ''Review of top cited HEP articles'' by Michael Peskin \cite{11} published from
1992 to 2003. The comparison reveals a remarkable correspondence between the scientific themes selected
by Peskin as themes of current importance and the themes appearing in the evolution of the classification
yielded by EqRank.


\section{Conclusions}

Let us recup. We formulated the notion of natural dynamics of partitions. A reasonable
clustering algorithm applied to evolving graphs should yield partitions with natural 
dynamics. We defined a number of coefficients characterizing the closeness of real partition
dynamics to a natural one. We introduced reindexation, which allowed us to transform the
initial partition to a new one possessing better structure and dynamics. Lastly, we 
applied this constructuion to the partition yielded by EqRank applied to the hep-th citation
graph, and demonstrated that the outcome of this procedure is in a good agreement with
the description of the themes given by an expert.

This work was supported in part by RFBR grant no. 03-02-17047.

\appendix
\section{The Tables}

Here we explain the columns of Tables 1 and 2 (The two tables could be joined into a single table;
Division into two tables is only to fit each of them
into a page).

The first column keeps the number of papers in a theme; In the second column, we keep the pairs of key words
characterizing a theme (the pairs were generated automatically by the titles of the papers); The third column
contains the titles of the most cited papers of a theme; The fourth column keeps the years a theme was
formed; The fifth, the year at wich Peskin mentioned the corresponding papers in his ''Review...'', and the last
column keeps either a title of the Section of Peskin's ''Review...'', or an excerpt from it characterising the theme.

Let us comment on how the years a theme was created (column four) were determined. For a theme, we map it
to a theme in the preceding partition with $Map_2$ (see above for the definition of $Map_2$). If it is
a stable theme, we map it again back in time, and continue until we obtain a new theme in the therminology 
introduced above. The year when the theme appears as new is the first one of the two years in column four,
the second year is the earliest year the theme appears as a stable one.

\end{document}